\documentclass[twocolumn,amsmath,amssymb,nofootinbib,preprintnumbers,showpacs]{revtex4}

\usepackage[pdftex]{graphicx}
\usepackage{epstopdf}

\usepackage{amsmath}
\usepackage{amssymb}
\usepackage{subfigure}
\usepackage{hyperref}
\usepackage{url}
\usepackage{xcolor}
\usepackage{color}
\definecolor{amaranth}{rgb}{0.9, 0.17, 0.31}
\definecolor{purple(munsell)}{rgb}{0.62, 0.0, 0.77}
\definecolor{americanrose}{rgb}{1.0, 0.01, 0.24}
\definecolor{palatinateblue}{rgb}{0.15, 0.23, 0.89}
\definecolor{royalblue(web)}{rgb}{0.25, 0.41, 0.88}
\definecolor{hanpurple}{rgb}{0.32, 0.09, 0.98}
\definecolor{beaublue}{rgb}{0.74, 0.83, 0.9}
\definecolor{carminered}{rgb}{1.0, 0.0, 0.22}
\definecolor{brightpink}{rgb}{1.0, 0.0, 0.5}
\hypersetup{ linktoc=all,
    colorlinks, linkcolor={palatinateblue},
    citecolor={brightpink}, urlcolor={palatinateblue}
}

\def\sideremark#1{\ifvmode\leavevmode\fi\vadjust{\vbox to0pt{\vss
 \hbox to 0pt{\hskip\hsize\hskip1em
 \vbox{\hsize2cm\tiny\raggedright\pretolerance10000
 \noindent #1\hfill}\hss}\vbox to8pt{\vfil}\vss}}}%
\begin{document}
   \title{Gravitational Synchrotron Radiation from Storage Rings}
   \author{Pisin Chen$^{1,2}$\footnote{pisinchen@phys.ntu.edu.tw}
      }

   \affiliation{%
   ~\\
$^{1}$Leung Center for Cosmology and Particle Astrophysics \& Department of Physics and Graduate Institute of Astrophysics, National Taiwan University, Taipei 10617, Taiwan\\
$^{2}$Kavli Institute for Particle Astrophysics and Cosmology, SLAC National Accelerator Laboratory, Stanford University, CA 94305, U.S.A.\\
}%

\pacs{}

\begin{abstract}
We reinvestigate the gravitational waves (GWs) induced by charged particles in storage rings. There are two major components in such GWs. One is the gravitational synchrotron radiation (GSR), i.e., the direct emission by the bending of the trajectory of a relativistic charged particle, much like the conventional electromagnetic synchrotron radiation (EMSR), albeit with characteristic difference in their radiation spectra. While the conventional EMSR spectrum peaks at the critical frequency, $\omega_c=\gamma^3\omega_0\gg\omega_0$, that of GSR peaks at the storage ring fundamental frequency $\omega_0$, which is much lower. The other component is the resonant conversion of EMSR to GWs at the same frequency through the coupling with fields of the storage ring bending magnets, i.e., the Gertsenshtein effect. Invoking LHC at CERN as a numerical example, we found that the spacetime perturbation associated with GSR, $h\sim 5\times 10^{-40}$, falls far below the sensitivity of GW detectors based on the LIGO-type Michelson interferometry approach. On the other hand, the GWs induced by the resonant conversion of EMSR have much higher frequencies and are thus localized. It is thus conceivable to detect them through the reverse conversion, the so-called `light shining through a wall', process.
\end{abstract}

\maketitle

\section{Introduction}\label{S1}   
LIGO's historical discovery of gravitational waves (GWs) \cite{LIGO: 2017} has ushered in a new era of gravity and astrophysics research. So far the cosmogenic GWs tend to be in the long wavelength regime, thus truly classical in the general relativity sense. One question naturally arises: Is it ever possible to have anthropogenic GW sources that can produce detectable signals in the short wavelength regime, or even particle-like gravitons? In fact, the possibility of GWs emitted from high energy particle storage rings has been investigated decades ago \cite{PG:1962,Khalilov:1972,KS:1974,Ritus:1990,Fargion:1988,Chen:1994}, but perhaps it was too much ahead of  its time, this pursuit was not further developed. In the aftermath of the LIGO discovery, the world is awakened to search for GWs with a wide range of frequencies. The time appears to be finally ripe to reevaluate the possibility of GWs radiated from storage rings. 

The revival of this author's interest in the topic was due largely to the recent ARIES APEC Workshop on Storage Rings and Gravitational Waves (SRGW2021) organized by F. Zimmermann in 2021 \cite{Zimmermann:2021}, where the main focus was on the possibility of using high energy particle storage rings as a detector for GWs from outer space. 

At an energy scale much lower than the Planck scale, the Einstein equation can be linearized as
\begin{equation}
\square\;\psi_{\mu\nu}=\frac{16\pi G}{c^4} T_{\mu\nu}\quad,\label{1}
\end{equation}
where $\psi_{\mu\nu}=h_{\mu\nu}-\eta_{\mu\nu}h/2$ is the trace-reversed metric perturbation around the flat spacetime $\eta_{\mu\nu}$ with the curved metric $g_{\mu}=\eta_{\mu\nu}+h_{\mu\nu}$, $\eta=\text{diag}(1,-1,-1,-1)$, and $T_{\mu\nu}$ is the energy-momentum stress tensor. Clearly, this equation provides solutions as propagating waves, i.e., the gravitational waves (GWs), with $T_{\mu\nu}$ serving as the source. 

We all know that the change of the quadrupole moment (in time) of a massive object can give rise to a GW. It was pointed out by Gertsenshtein \cite{Gertsenshtein:1962} that when a propagating EM wave traverses a transverse background EM field, there is a nontrivial stress tensor, $T^{f}_{\mu\nu}$, which can \textit{resonantly excite} a GW with the same frequency as the initial propagating EM wave. Specifically, 
\begin{equation}
T^{f}\sim\big(F^{em}+F^0\big)\big(F^{em}+F^0\big)\quad,
\end{equation}
where $F^{em}$ and $F^0$ are the Maxwell field tensor of the propagating and the background EM field, respectively. It should be noted that the square of the background field, $F^0F^0$, bears no relation to the motion of the particle and we shall ignore it in the following. There is also no need to consider the square of the propagating EM field, $F^{em}F^{em}$, since almost everywhere $F^{em} \ll F^0$ except at small distance from the particle. But this has been taken into account in the mass renormalization and has been included in $T^p$. Therefore, in effect only the quadratic terms contribute to the GW generation and so we identify $T^f=F^{em}F^0+F^0F^{em}$ from here on.

In the case of GWs emitted by relativistic charged particles in a storage ring, both the massive particle itself and the conventional electromagnetic synchrotron radiation (EMSR) emitted by the same particle contribute to the source for GWs, i.e., 
\begin{equation}
T_{\mu\nu}=T^{p}_{\mu\nu}+T^{f}_{\mu\nu}\quad.
\end{equation}
These two source terms are independent of each other, and we can treat the GWs induced by them separately. For the part of the GW that is generated by $T^{p}_{\mu\nu}$, the electromagnetic interaction serves only as a means to bend the particle trajectory and the mass of the particle acts just like a gravitational "charge". For this reason we shall call this subset of the GW the \textit{gravitational synchrotron radiation} (GSR). 

In this paper, we will first derive the GSR and then turn to the resonant conversation of EMSR to GWs. We will then take LHC as an example to evaluate the possible yields of GWs. 

\section{Gravitational Synchrotron Radiation}

As a general property of a wave equation, in the wave zone we have, from Eq.~\eqref{1},
\begin{equation}
\psi_{\mu\nu}(\vec{R})=\frac{4G}{R}\int d\omega e^{-i\omega(t-R)}T^p_{\mu\nu}(\omega, \vec{k})\quad,\label{2}
\end{equation}
where $\vec{R}=R\hat{n}$ and $R$ is the distance to the observation point. The momentum of the gravitational radiation is $k=\omega$ and $\vec{k}=\omega\hat{n}$. For convenience, we set $c=1$. Figure 1 shows the relevant parameters involved in GSR.

\begin{figure}[!htb]
\minipage{0.4\textwidth}
  \includegraphics[width=\linewidth]{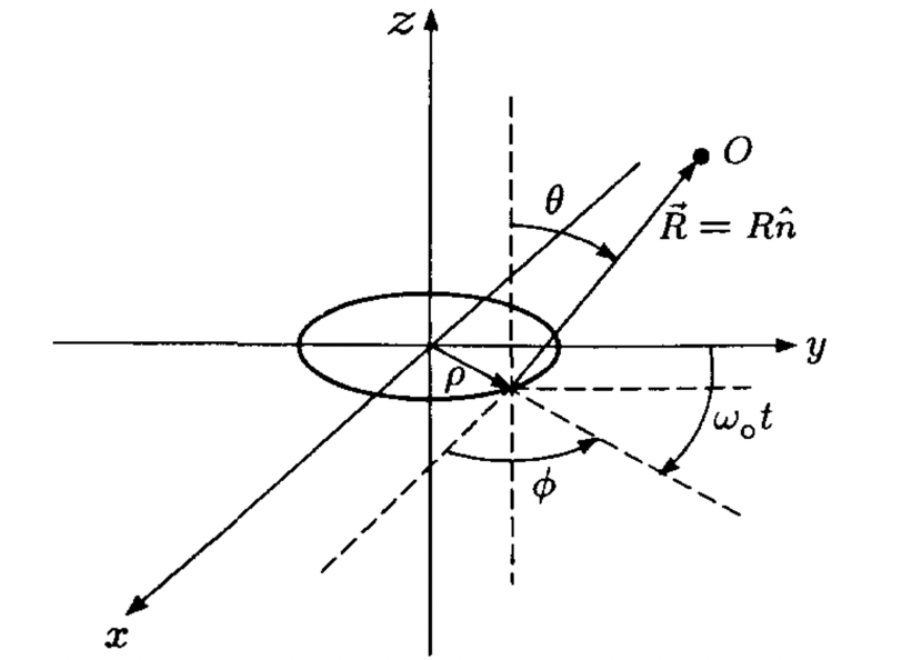}
  \caption{Coordinates involved in the gravitational synchrotron radiation.}\label{F1}
\endminipage\hfill
\end{figure}

For GSR, it can be shown in the momentum space that the component of $T^p_{\mu\nu}$ that is doubly transverse to the tangent of the circular orbit contributes to the radiation. Replacing $\int d\omega$ with $\omega_0\sum_n$, we have 
\begin{widetext}
\begin{equation}
\begin{aligned}
T^{p}_{_{\perp\perp} n}(k)&=2\pi\gamma m\rho^2\omega_0 e^{in(\phi-\pi/2)}
\left\{\sin^2\phi J_{n}(\xi)+2i\sin\phi\cos\phi\left[-\frac{n}{\xi^2}J_{n}(\xi)+\frac{n}{\xi}J'_{n}(\xi) \right]-\cos{2\phi}J''_{n}(\xi) \right\},\label{3}
\end{aligned}
\end{equation}
\end{widetext}
dominates the contribution. Here $n$ is the harmonic number, $\xi=k\rho\beta\sin\theta=n\beta c\sin\theta$, and $\omega_0=c/\rho$ is the orbital frequency, $\theta,\phi$ are the polar coordinates defined in Fig.~\ref{F1}, and $J_n$ is the Bessel function of the first kind.  
The GSR radiation power is then \cite{Landau-Lifshitz} 
\begin{align}
W_{G}(\omega)&=\frac{R^2}{32\pi G}\int d\Omega\left[\partial_{0}\psi^{\mu\nu}\partial_{i}\psi_{\mu\nu}-\frac{1}{2}\partial_{0}\psi^{\mu}_{\;\mu}\partial_{i}\psi^{\nu}_{\;\nu} \right]n^{i}\nonumber
\\
&=\frac{G\omega^2}{2\pi}\int d\Omega\;T^{\mu\nu}(\vec{k})T_{\mu\nu}(-\vec{k})\quad,\label{4}
\end{align}
where $n_i$ is the $i$th component of $\vec{n}$. 
Inserting Eq.~\eqref{3} into Eq.~\eqref{4} we obtain the GSR power spectrum
\begin{equation}
\begin{aligned}
\frac{dW_{GSR}}{dx}&=\frac{3\sqrt{\pi}}{32}\frac{Gm^2\gamma^4\omega_0^2}{c}
\\
&\times\left[3x^{-1/3}\Phi(y)-5x^{1/3}\Phi'(y)+3x\Phi_2(y) \right],
\label{5}
\end{aligned}
\end{equation}
where $x=\omega/\omega_c,y=x^{2/3},\omega_c=\gamma^3\omega_0$ is the critical frequency of the synchrotron radiation, $\Phi$ is the Airy function, and
\begin{equation}
\Phi_2(y)=\frac{y^{1/2}}{2^{2/3}\pi^{1/2}}\int_{-\infty}^{\infty}dz\Phi^2(y(1+z^2)/2^{2/3})\quad.\label{6}
\end{equation}

\begin{figure}[!htb]
\minipage{0.5\textwidth}
  \includegraphics[width=\linewidth]{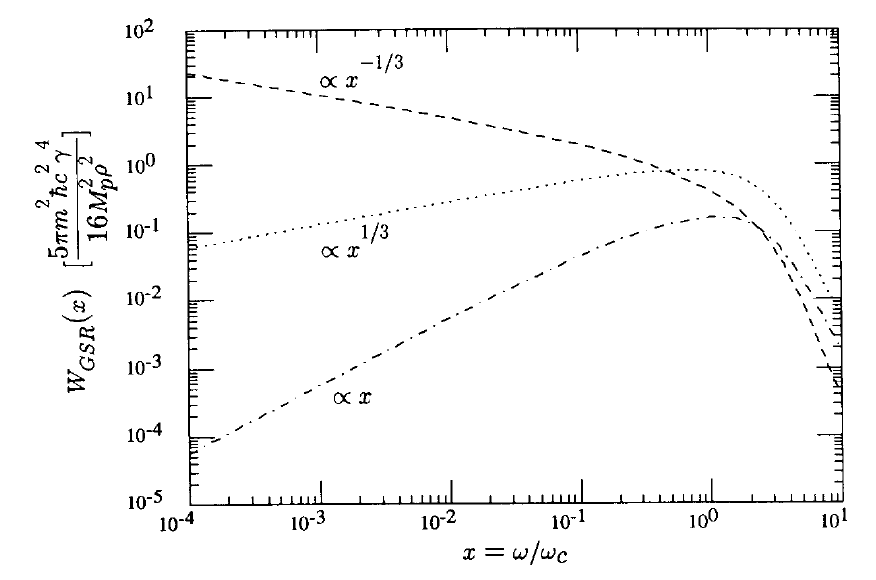}
  \caption{Gravitational synchrotron radiation power spectrum.}\label{F2}
\endminipage\hfill
\end{figure}

Figure~\ref{F2} shows the GSR spectrum with the contribution from the three terms in Eq.~\eqref{5} plotted separately. At small $x$, the spectrum scales as $x^{-1/3},x^{1/3}$, and $x$, respectively (See Fig.~\ref{F2}). 

\section{Estimated GSR Graviton Yields}

Further integrating over the spectrum, we find the total power
\begin{equation}
W_{GSR}=\frac{5\pi}{16}\frac{Gm^2c\gamma^4}{\rho^2}=\frac{5\pi}{16}\frac{m^2}{m_P^2}\frac{\hbar c^2\gamma^4}{\rho^2}\quad,\label{7}
\end{equation}
where $m_P\equiv (\hbar c/G)^{1/2}$ is the Planck mass. Although the total power scales as $\gamma^4$, same as that for the electromagnetic synchrotron radiation (EMSR), the GSR power is dominated instead by the fundamental frequency, $\omega_0=c/\rho$, due to its scaling law $x^{-1/3}$ in the small $x$ limit (See Fig.~\ref{F2}). This is characteristically different from that in the EMSR, where the dominant frequency is $\omega_c=\gamma^3\omega_0\gg \omega_0$. Therefore not only all $N$ particles in a particle bunch in a storage ring radiate GSR coherently, all the $n_b$ bunches in the ring can radiate coherently so long as the bunches are not distributed symmetrically around the ring. The total rate of graviton emission is then
\begin{equation}
N_{GSR}\sim 5.6n_b^2N^2\frac{m^2}{m_P^2}\frac{c\gamma^4}{\rho}.\label{8}
\end{equation}
Note, however, that this is the total yield around the ring. The collectable signal is much reduced if the detector is concentrated at a single location with a finite solid angle. Furthermore, at such low (fundamental) frequencies the notion of gravitons as discrete entities in the GW is questionable. We remind again that this is only a fraction of the total graviton yield from such an electromagnetic system, where the EMSR can also convert into gravitons through resonant conversion. We will return to this issue at the end of the next section.


\section{Resonant Conversion of Electromagnetic Synchrotron Radiation}\label{S3}

We now turn to the resonant conversion of electromagnetic synchrotron radiation into the gravitational radiation, which is induced by the field tensor $T_f$ described in Eq.(2).
Earlier, this effect was invoked in the investigation of GW production from high energy storage rings \cite{4}. it has been shown that \cite{7}
\begin{equation}
W_G(\omega)=\frac{\pi}{4}\frac{1}{\alpha}\frac{m^2}{m_P^2}\left(\frac{L}{\lambda_c}\frac{B}{B_c}\right)^2\left[1-\frac{\sin(\omega L)}{\omega L} \right]^2W_{EM}(\omega),\label{9}
\end{equation}
where $\lambda_c$ is the Compton wavelength, $B_c\equiv m^2c^3/e\hbar\sim 4.4\times 10^{13}$ Gauss is the Schwinger critical field, and $W_{EM}$ is the power spectrum of the propagating EM field, in our case the EMSR emitted by the same charged particle in the storage ring. The square represents the factor from the Fourier Spectrum of the background field. We see that this form factor is essentially of order unity for wavelengths $\lambda\lesssim 2L$, where the last zero at $\sin(2\pi L/\lambda)=\sin\pi$ occurs. Beyond this wavelength the GWs are largely suppressed. 

We see that in addition to the GSR there is also a resonant conversion with the rate 
\begin{equation}
N_{res}\sim\frac{1}{\alpha}\frac{m^2}{m_P^2}\Big(\frac{L}{\lambda_c}\frac{B}{B_c}\Big)^2 N_{\gamma}, 
\end{equation}
where $N_{\gamma}$ is the EMSR photon number. Since EMSR is dominated by the critical frequency $\omega_c=\gamma^3\omega_0\gg\omega_0$, this radiation is not coherent in high energy storage rings that we considered. As a result, the relative yield is $N_{res}/N_{GSR}\sim (1/n_bN)(B/B_c)^2(L/\lambda_c)^2$. Take, for the sake of discussion, $B\sim 10$Tesla and $L\sim 10$m. Then since the number of particles per bunch is of the order $10^{11}$ in 
LHC, the relative yield is reduced by roughly a factor of $1/10n_b$. 

Let us take LHC as an example. With proton energy at 7TeV, $N=10^{11}$, $n_b=3000$, there will be of the order $N_{GSR}\sim 10^{10}$ gravitons emitted through gravitational synchrotron radiation at the fundamental frequency $\omega_0\simeq 70 {\rm kHz}$, and of the order $N_{res}\sim 10^5$ gravitons emitted through resonance conversion at the critical frequency $\omega_0\gamma^3\simeq 5\times 10^4 {\rm GHz}$ per year. With in mind the extreme weakness of the gravitational interaction, these numbers are modest in terms of the detectability, to say the least.

\section{GSR Induced Spacetime Metric Perturbations}
Conventional approach to the GW detection such as LIGO measures the perturbations of the spacetime metric from that of the flat spacetime, $h=\sqrt{h_{\mu\nu}h^{\mu\nu}}$. We wish to determine the magnitude of the perturbation $h$ associated with SGR from storage rings. Approximately, 
\begin{equation}
h\sim \vert h_{\mu\nu}-\eta_{\mu\nu}h/2 \vert \sim \vert\psi_{\mu\nu}\vert.
\end{equation}
Note that in our convention, the dimensionality of the relevant fundamental constants and variables are 
\begin{equation}
[G]=2, [\hbar]=1, [\psi_{\mu\nu}(x)]=0, [T_{\mu\nu}(x)]=2, [T_{\mu\nu}(k)]=0,  
\end{equation}
The Fourier transformation of $\psi(R)$ in the momentum space is 
\begin{equation}
\begin{aligned}
\psi^p_{_{\perp\perp}}(\omega)=\frac{4G\omega_0}{R}\sum_{n=1} e^{-in\omega_0 t}e^{in(\phi-\pi/2+R)}T^p_{_{\perp\perp}n}(\vec{k}).
\end{aligned}
\end{equation}
As discussed before, at low frequencies the GSR is dominated by the first term of Eq.(5) that scales as $x^{-1/3}$. Thus
\begin{equation}
T^{p}_{_{\perp\perp}}(k)\simeq 2\pi\gamma m\rho^2\omega_0\sum_{n=1}^{\infty} e^{in(\phi-\pi/2)}\sin^2\phi J_{n}(\xi).  
\end{equation}
Let us set $\phi=\pi/2$. Since $\sum_{n=1}^{\infty}J_n(x) \sim\mathcal{O}(1)$, the metric perturbation induced by a single orbiting particle is 
\begin{equation}
\hat{h}\sim \vert \psi_{\mu\nu}\vert \sim \gamma (\rho\omega_0)^2\frac{m}{m_{P}}\frac{l_P}{R},
\end{equation}
where $l_P=1/m_P$ is the Planck length. Note that the fundamental mode is related to the radius of the storage ring by $\rho\omega_0=c=1$. Typically in a high energy particle storage ring there are $N$ particles in a cluster (a bunch), and a train of $n_b$ bunches orbiting closely together. So the total $h$ is 
\begin{equation}
h\sim \sum \hat{h} \sim \gamma (\rho\omega_0)^2\frac{m}{m_{P}}\frac{l_P}{R}n_bN.
\end{equation}

Let us again take LHC as an example. The radius of the LHC ring is $\rho=4300{\rm m}$. The proton energy is 7GeV. So $\gamma\sim 7500$. With $N=10^{11}$ and $n_b=3000$, and tasking the shortest distance for detection, $R\sim \rho$, we find $h\sim 5\times 10^{-40}$.

\section{Discussion}\label{S8}
In this paper we reinvestigate the gravitational synchrotron radiation induced by high energy charged particles in storage rings.
Invoking the existing LHC at CERN as an example, we found that $h\sim 5\times 10^{-40}$, which is far below the sensitivity of GW detectors based on the LIGO-type Michelson interferometry approach. On the other hand, the GWs induced from the resonant conversion of EMSR, with the critical frequency $\omega_c=\gamma^3\omega_0\gg\omega_0$, would be much more localized and therefore behave more particle-like than wave-like, that is, more like a graviton. 

Perhaps this particle-like character would inspire a different detection method that is more advantageous. For example, one may invoke the reverse process of resonant conversion with a long magnet to convert gravitons back to photons at the same EMSR critical frequency as a means of detection \cite{7,Chen:1994}. Indeed, the same philosophy has also been applied to the search for axions, often dubbed as {\it light shining through a wall} \cite{Redondo:2011}, such as the OSQAR experiment at CERN \cite{OSQAR}.

\acknowledgments
I appreciate helpful discussions with G. Diambrini-Palazzi, J. Ellis, D. Fargion, P. Huet, C. Pellegrini, and G. Veneziano during the \textit{First International Conference on Phenomenology of Unification from Present to Future}, Rome, Italy, 1994, where my original work was presented. The revival of my interest in this topic owes largely to the very interesting ARIES APEC online workshop on Storage Rings and Gravitational Waves (SRGW2021) organized by J. Ellis and F. Zimmermann, in 2021 \cite{Zimmermann:2021}. I appreciate helpful discussion with the workshop participants, in particular R. T. D'Agnolo, J. Ellis, J. Jowett, K. Oide, and F. Zimmermann. I appreciate various helpful discussions with my student Kuan-Nan Lin, who also checked carefully through my calculations. 

\appendix


\begin{thebibliography}{99}

\bibitem{LIGO: 2017}
B. P. Abbott et al., Phys. Rev. Lett. {\bf 116}, 061102 (2016).

\bibitem{PG:1962} 
V. I. Pustovoit and M. E. Gertsenshtein, Sov. Phys. JETP {\bf 15}, 116 (1962). 

\bibitem{Khalilov:1972}
V. R. Khalilov, J. M. Loskutov, A. A. Sokolov, and I. M. Ternov, Phys. Lett. {\bf 42A} , 43(1972).

\bibitem{KS:1974}
I. B. Khriptovich and E. V. Shuryak, Sov. Phys. JETP {\bf 38}, 1067 (1974). 

\bibitem{Ritus:1990}
A. I. Nikishov and V. I. Ritus, Sov. Phys. JETP {\bf 69}, 876 (1990); A. I. Nikishov and V. I. Ritus, Sov. Phys. JETP {\bf 71}, 643 (1990).

\bibitem{Fargion:1988}
G. Diambrini-Palazzi and D. Fargion, Phys. Lett. {\bf B197}, 302 (1987); G. Diambrini-Palazzi, "On the Production and Detection of Gravitational Waves from Artificial Sources", CERN EP-88/112, 1988; D. Fargion, "Fourier Transform of Gravitational Synchrotron Radiation", Univ. Rome Preprint N.588, 1988. 

\bibitem{Chen:1994}
P. Chen, ``Resonant Photon-Graviton Conversion in EM Fields: from Earth to Heaven", in \textit{First International Conference on Phenomenology of Unification from Present to Future}, Rome, Italy, 1994 (World Scientific); SLAC-PUB-6666 (1994).

\bibitem{Zimmermann:2021}
A. Berlin, M. Br\"uggen, O. Buchmueller, P. Chen, R. T. D'Agnolo, R. Deng, J. R. Ellis, S. Ellis, G. Franchetti, A. Ivanov, J. M. Jowett, A. P. Kobushkin, S. Y. Lee, J. Liske, K. Oide, S. Rao, J. Wenninger, M. Wellenzohn, M. Zanetti, F. Zimmermann, {\it Storage Rings and Gravitational Waves: Summary and Outlook}, ed. J. Ellis, F. Zimmermann, arXiv:2105.00992 (2021).

\bibitem{Gertsenshtein:1962}
M. E. Gertsenshtein, Sov. Phys. JETP {\bf 14}, 84 (1962).

\bibitem{Landau-Lifshitz}
Landau and Lifshitz, \textit{Classical Theory of Fields}, Pergamon Press (1968).

\bibitem{4} 
G. Diambrini-Palazzi, Part. Accel. {\bf 33}, 195 (1990).

\bibitem{7}
P. Chen, Mod. Phys. Lett. {\bf 6}, 1069 (1991). 

\bibitem{Redondo:2011}
J. Redondo, A. Ringwald, Cont. Phys. {\bf 52}, 211 (2011).

\bibitem{OSQAR}
P. Pugnat et al., OSQAR Annual Report 2020, CERN-SPSC-2020-025/SPSC-SR-280 (2020). 





 

 

     




    
\end{thebibliography}
\end{document}